\pdfoutput=1
\documentclass{article}
\usepackage{ijcai19}

\usepackage{times}
\usepackage{soul}
\usepackage[utf8]{inputenc}
\usepackage[small]{caption}
\usepackage{amsfonts}
\usepackage{graphicx}
\usepackage{amsmath}
\usepackage{booktabs}
\usepackage{algorithm}
\usepackage{algorithmic}
\usepackage{color}
\usepackage{multirow}
\usepackage{subcaption}
\usepackage{mathtools, bm}
\usepackage{amsmath}
\usepackage{breqn}
\usepackage{cleveref}

\newcommand{\tool}{\textbf{MatchGNet}~}
\newcommand{\toole}{\textbf{MatchGNet}}

\newcommand{\nop}[1]{}
\newcommand{\EventIn}[3]{}
\newcommand{\apt}[3]{Advanced Persistent Threat}

\title{Heterogeneous Graph Matching Networks}

\author{
Shen Wang$^{1,2,}$\thanks{Work done during an internship at NEC Labs America.}$^,$\thanks{The first two authors contributed equally.}\and
Zhengzhang Chen$^{2,}$\footnotemark[2]\and
Xiao Yu$^2$\and
Ding Li$^2$\and
Jingchao Ni$^2$\and
Lu-An Tang$^2$\and\\
Jiaping Gui$^2$\and
Zhichun Li$^2$\and
Haifeng Chen$^2$ \And
Philip S. Yu$^{1,3}$
\affiliations
$^1$University of Illinois at Chicago, USA\\
$^2$NEC Laboratories America, USA\\
$^3$Tsinghua University, China\\
\emails
\{swang224, psyu\}@uic.edu,
\{zchen, xiao, dingli, jni, ltang, jgui, zhichun, haifeng\}@nec-labs.com
}

\begin{document}

\maketitle

\begin{abstract}
 Information systems have widely been the target of malware attacks. Traditional signature-based malicious program detection algorithms can only detect known malware and are prone to evasion techniques such as binary obfuscation, while behavior-based approaches highly rely on the malware training samples and incur prohibitively high training cost. To address the limitations of existing techniques, we propose \toole,\nop{an attentional heterogeneous Graph Neural Network based model to learn the distinguishable representations of program based on the invariant graph modeling of the program's execution behavior.}
 a heterogeneous Graph Matching Network model to learn the graph representation and similarity metric simultaneously based on the invariant graph modeling of the program's execution behaviors.\nop{learn the distinguishable representation of the invariant program graph by modeling the program's execution behaviors and perform effective graph similarity matching between a pair of program graphs.}\nop{joint learn the distinguishable representation and similarity metric for program graph on the invariant graph modeling of the program's execution behavior.} We conduct a systematic evaluation of our model and show that it is accurate in detecting malicious program behavior and can help detect malware attacks with less false positives. \tool outperforms the state-of-the-art algorithms in malware detection by generating 50\% less false positives while keeping zero false negatives.

\end{abstract}

\maketitle

\section{Introduction}
\label{sec:intro}

With information systems playing ubiquitous and indispensable roles in many modern industries including financial services and retail businesses, cyber security undoubtedly bears the utmost importance in the daily system management. Under constant cyber-attacks happening every day, incidents causing significant financial losses and leakage of sensitive customer data appear in media headlines every now and then. 
Among these attacks, malware/malicious program attacks are the most widespread and costly type, and they are growing fast to target at more companies and organizations. According to a study report published by Accenture\footnote{www.accenture.com/us-en/insight-cost-of-cybercrime-2017}, 98\% of the participating companies experienced malware attacks in 2016 and 2017, and these attacks cost companies an average of \$2.4 million. Symantec also finds that the number of groups using destructive malware was up by 25\% in 2018\footnote{www.symantec.com/security-center/threat-report}. With imminent malware attacks, protecting key information systems has become the top priority for many companies and organizations.

Malware/malicious program detection as the first line of defense against attacks mainly uses two types of approaches: signature-based and behavior-based. Signature-based approaches~\cite{Damodaran2017,David2017} check program signatures against a known malware database. Such signatures can be instruction sequences, system calls, loaded dynamic libraries, and so on. While signature-based approaches are widely adopted by antivirus software due to their time efficiency, they are mostly limited to detect known malware and prone to evasion techniques such as binary obfuscation. Nowadays, such signature-based approaches are facing difficulties from zero-day malware attacks, which are highly polymorphic and rarely have known signatures~\cite{Damodaran2017}.

Behavior-based approaches~\cite{Rieck2008,Domagoj2011,Jha2013,Bernardi2018,Saracino2018} as an improvement over the signature-based ones learn the program behaviors in terms of dependencies between system-elements (\textit{i.e.}, system-calls or API calls), and classify a program to either the malicious or the benign category using the learned model. The program behaviors captured by these approaches generally reflect real program intentions, thereby complementing static signatures that can easily be manipulated. On the other hand, behavior-based approaches incur prohibitively high training cost. In order to learn malicious program behavior, one must collect enough malware samples, and then execute them repeatedly in a controlled environment to observe their runtime behaviors. Insufficient sampling and observations of malicious behavior can unavoidably limit the detection capability, and malware attacks can also evolve via adversarial learning to evade detection.

In contrast to those malware detection algorithms focusing on detecting \textit{known} malware, the goal of this paper is to design an effective  \textit{data-driven} approach for detecting \textit{unknown} malicious programs. We define a malicious program as a process that is unknown to the execution environment and behaves differently to all existing benign programs. Slightly more formally, given a target program with corresponding process event data (\textit{e.g.}, a program opens a file or connects to a server) during a time window, we perform similarity learning and check whether the behavior of the program is similar to that of any existing benign programs. If some highly similar programs exist, the model outputs \textit{top-k} most similar programs with their IDs/names and similarity scores\nop{predicted label should be $-1$ and the \textit{top-k} most similar program IDs/names would be recommended}; otherwise, it triggers an alert.

It is difficult to detect unknown malicious programs due to four major challenges: (1) The nonlinear and hierarchical heterogeneous relations/dependencies among system entities. The operations made by programs have nonlinear and hierarchical heterogeneous dependencies, and they work together in a highly complex and coordinated manner. Simple methods that neglect these dependencies may still yield high false positive rates; (2) Lack of intrinsic distance/similarity metrics~\cite{zhang2015categorical} between two programs. Consider that in real computer systems, given two programs with thousands of system events related to them, it is a non-trivial task to measure their distance/similarity based on the categorical event data; (3) The exponentially large event space. An information system typically deals with a large volume of system event data (normally more than $10,000$ events per host per second)~\cite{Dong:2017}; (4) Aliases of programs. Different versions or updates of the same program have different signatures and may also have different executable names. Thus, a simple method that keeps a white-list of program IDs/signatures could be problematic. 

To address these challenges, we propose \toole, a data-driven graph \underline{match}ing framework to learn the program representation and similarity metric via \underline{G}raph Neural \underline{Net}work. In particular, we first design an invariant graph modeling to capture the heterogeneous interactions/dependencies between different pairs of system entities. To learn the program representation from the constructed heterogeneous invariant graph, we propose a hierarchical attentional graph neural encoder. Finally, we propose a similarity learning model via Siamese Network to train the parameters and perform similarity scoring between an unknown program and the existing benign programs. 
\tool trains the model on existing benign programs, not on any malware/malicious program samples. As a result, it can identify an unknown malicious program whose behavioral representation is significantly different to any of the existing benign programs by similarity matching. 
We conduct an extensive set of experiments on real-world system provenance data to evaluate the performance of our model. The results demonstrate that \tool can accurately detect malicious programs. In particular, it can detect fake and unknown programs with an average accuracy of 97\% and 96\%, respectively. We also apply \tool to detect realistic malware attacks. The results show that it can reduce false positives of the state-of-the-art by 50\%, while keeping zero false negative.

\section{Motivating Example and Related Work}
\label{sec:motiv}
\begin{figure}[t!]
  \centering
  \includegraphics[width=0.48\textwidth]{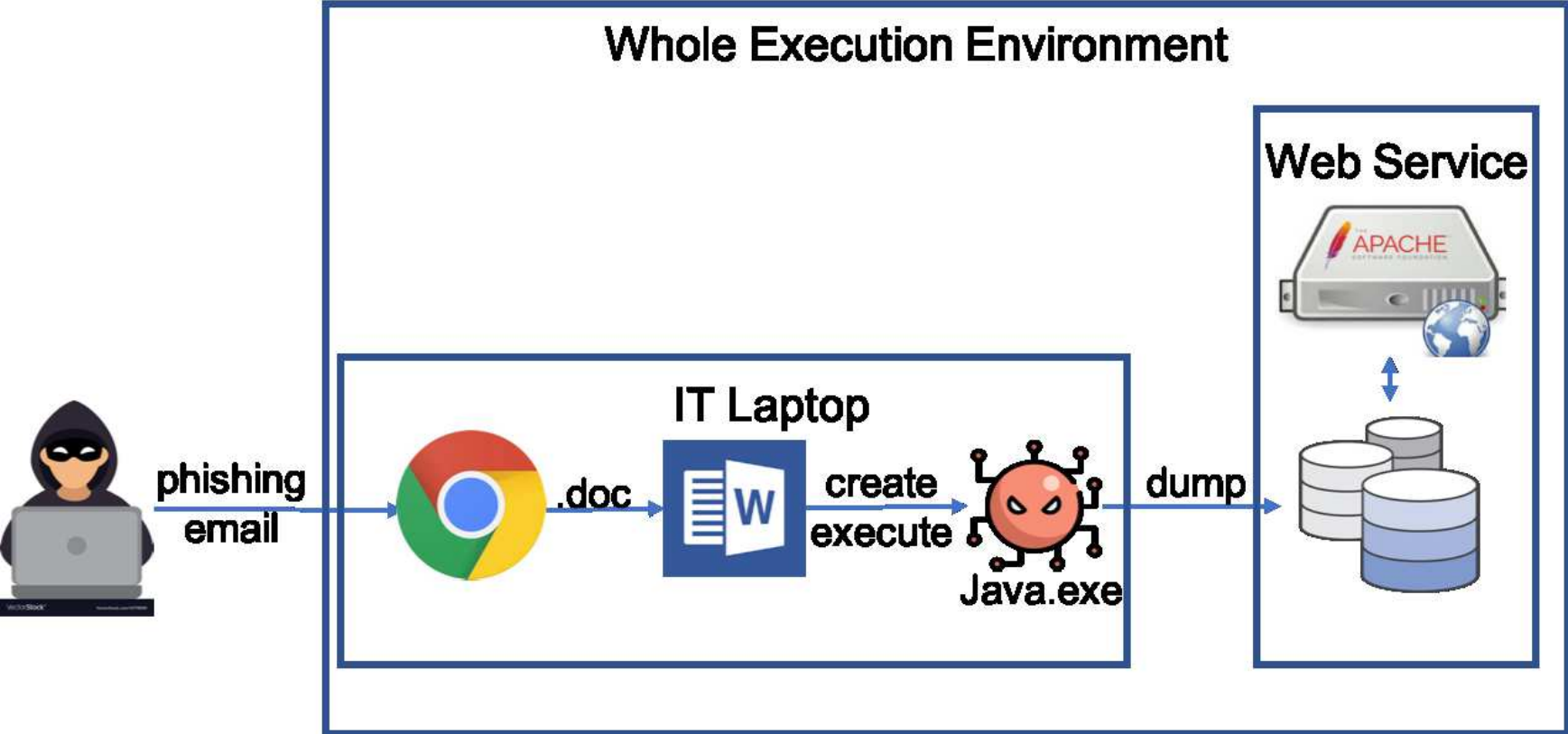}
\caption{An example of phishing email attack.}
\label{fig:phishingemail}
\end{figure}

In a phishing email attack as shown in Figure~\ref{fig:phishingemail}, to steal sensitive data from the database of a computer/server, the adversary exploits a known venerability of Microsoft Office\footnote{https://
nvd.nist.gov/vuln/detail/CVE-2008-0081} by sending a phishing email attached with a malicious .doc file to one of the IT staff of the enterprise. When the IT staff member opens the attached .doc file through the browser, a piece of malicious macro is triggered. This malicious macro creates and executes a malware executable, which pretends to be an open source Java runtime (Java.exe). This malware then opens a backdoor to the adversary, subsequently allowing the adversary to read and dump data from the target database via the affected computer. 

\begin{figure}[htp]
  \centering
  \includegraphics[width=0.48\textwidth]{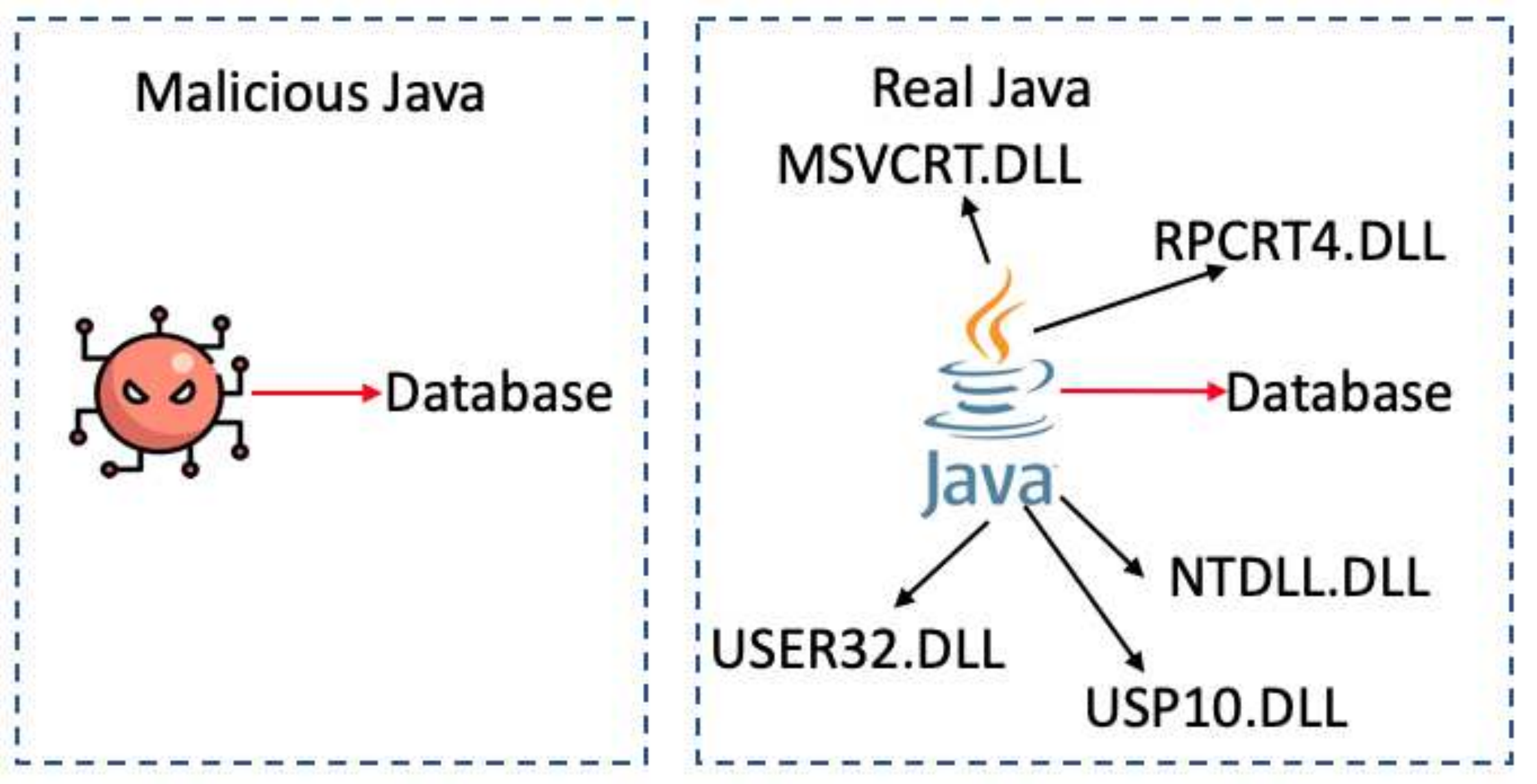}
\caption{Malicious Java.exe vs. normal Java runtime. }
\label{fig:java}
\end{figure}

Signature-based or behavior-based malware detection approaches generally do not work well in detecting the malicious program in our example. As the adversary can make the malicious program from scratch with binary obfuscation, signature-based approaches \cite{David2017,Bernardi2018} would fail due to the lack of known malicious signatures. Behavior-based approaches~\cite{Domagoj2011,Jha2013,Bernardi2018} may not be effective either, unless the malware sample has previously been used to train the detection model.

It might be possible to detect the malicious program using existing host-level anomaly detection techniques~\cite{lin2012intelligent,jyothsna2011review,Chen:2016,Cao:2018:BCD:3269206.3272022}.
These host-based anomaly detection methods can locally extract patterns from process events as the discriminators of abnormal behavior. 
However, such detection is based on observations of single operations, and it sacrifices the false positive rate to detect the malicious program~\cite{Lin:2018:CAR:3269206.3272013}.
For example, the host-level anomaly detection can detect the fake ``Java.exe'' by capturing the database read. However, a Java-based SQL client may also exhibit the same operation. If we simply detect the database read, we may also classify normal Java-based SQL clients as abnormal program instances and generate false positives. In the enterprise environment, too many false positives can lead to the alert fatigue problem~\cite{7931672,Lin:2018:CAR:3269206.3272013}, causing cyber-analysts to fail to catch up with attacks.

To accurately separate the database read of the malicious Java from the real Java instances, we need to consider the higher semantic-level context of the two Java instances. As shown in \Cref{fig:java}, the malicious Java is a very simple program and directly accesses the database. On the contrary, a real Java instance has to load a set of .DLL files in addition to the database read. By comparing the context of the fake Java instance, we can find that it is not normal and precisely report it as a malicious program instance. Thus, in this paper, we propose a Graph Neural Network based approach to learn the semantic-level context of program instances.

In recent years, Graph Neural Network (GNN) approaches~\cite{defferrard2016convolutional,hamilton2017inductive,velickovic2017graph,kipf2016semi,gilmer2017neural,wu2019comprehensive,wang2019adversarial,ying2018hierarchical,chen2018fastgcn}
have been proposed for graph-structured data. They try to accelerate convolution operations, reduce the computational cost, and extend the current graph convolution. The goal of GNN is to learn the representation of the graph, in the node level or graph level. Because of its remarkable graph representation learning ability, GNN has been explored in various real-world applications, such as
healthcare \cite{mao2019medgcn,WangSBDM}, chemistry \cite{fout2017protein,zitnik2018modeling}, 
and security system \cite{WangAHGNN}. 
\nop{However, majority of previous approaches focus on the classification problem for a homogeneous network. Compared to these approaches, our work is different in two folds: 1) our work focuses on learning from the heterogeneous graph, rather than the homogeneous graph; 2) our work learns the graph representation and similarity metric jointly, rather than independently mapping each graph to a vector.}  

In parallel to the Graph Neural Network, graph similarity matching has been studied extensively in the machine learning community \cite{yan2004graph,shasha2002algorithmics,willett1998chemical,raymond2002rascal}. The similarity metric can be categorized into two types: exact matching (isomorphism test) \cite{yan2004graph,shasha2002algorithmics} and structure similarity measures (graph edit distance) \cite{willett1998chemical,raymond2002rascal}. Besides, in the computer vision community, learning-based metrics \cite{weinberger2009distance,sun2014deep} are proposed to match the image data. They compute the similarity score with either hand-engineered features or hand-designed metrics. Compared to these approaches, our work is different in two folds: 1) we focus on learning from the heterogeneous graph, rather than the homogeneous graph; and 2) we learn the graph representation and similarity metric simultaneously.  

\section{Heterogeneous Graph Matching Networks}
\subsection{Overview}
\begin{figure*}[htb]
\centering
\includegraphics[width =0.7\linewidth]{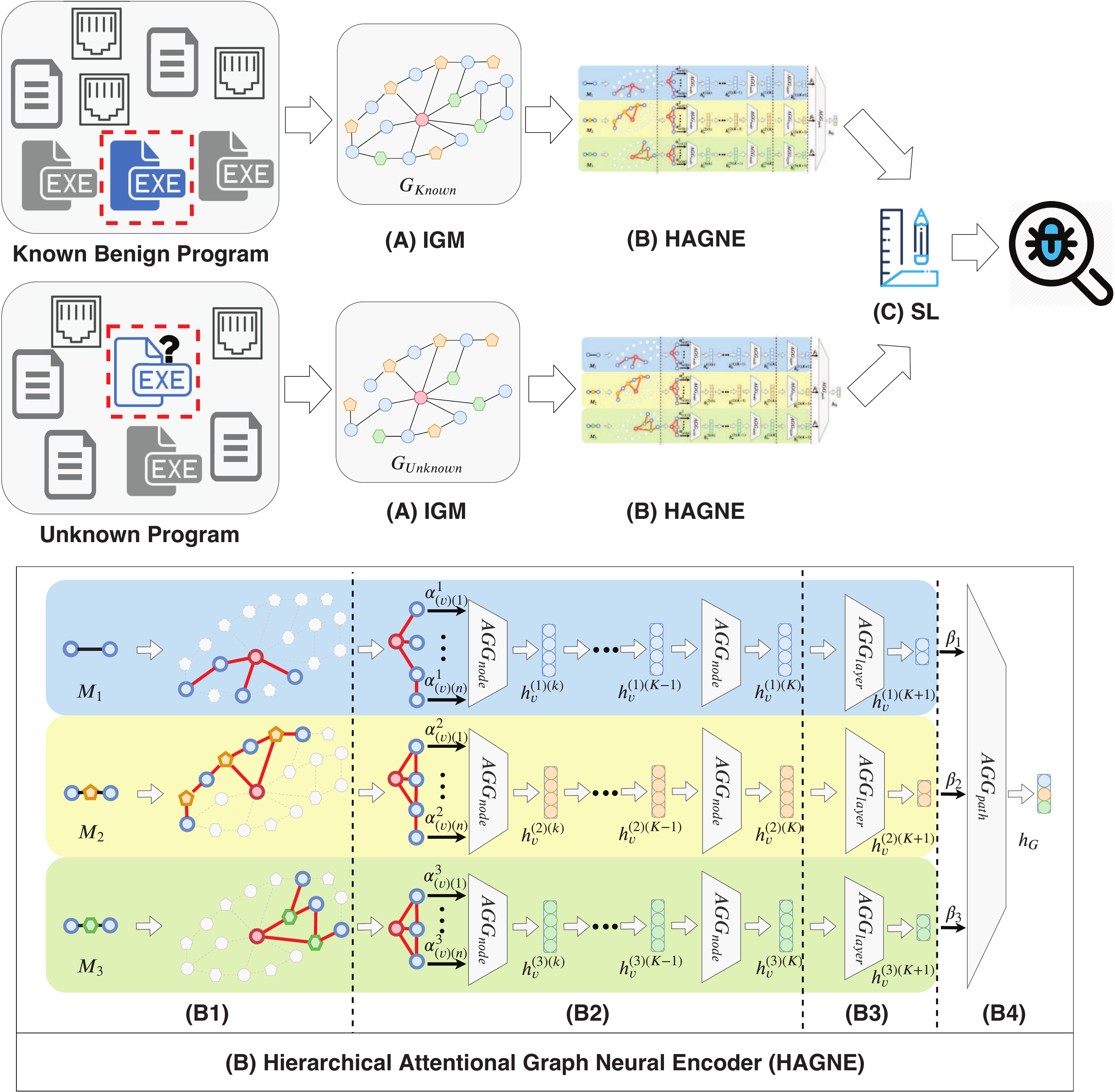}
\caption{The architecture of MatchGNet consists of three key modules: (A) Invariant Graph Modeling (IGM), (B) Hierarchical Attentional Graph Neural Encoder (HAGNE), and (C) Similarity Learning (SL). IGM models system event data as a heterogeneous invariant graph. HAGNE encodes the heterogeneous graph into an embedding by four components: (B1) Heterogeneity-aware Contextual Search, (B2) Node-wise Attentional Neural Aggregator, (B3) Layer-wise Dense-connected Neural Aggregator, and (B4) Path-wise Attentional Neural Aggregator. SL trains HAGNE to be more distinguishable between an unknown program and known benign programs and computes the similarity score for unknown malware detection.}
\label{fig:Architecture}
\end{figure*}

To address the challenges introduced in \Cref{sec:intro}, we propose a heterogeneous Graph Matching Network framework, \toole, with three key modules: Invariant Graph Modeling (\textbf{IGM} or Step (A)), Hierarchical Attentional Graph Neural Encoder (\textbf{HAGNE} or Step (B)), and Similarity Learning (\textbf{SL} or Step (C)) as illustrated in \Cref{fig:Architecture}. The \textbf{IGM} module models the system event data as a heterogeneous invariant graph, which can capture the program's behavior profile. Then, we formulate the malicious program detection as a heterogeneous graph matching problem and solve it with \textbf{HAGNE} and \textbf{SL}. 

Given two graphs $G_{(1)} = (V_{(1)},E_{(1)})$ and $G_{(2)} = (V_{(2)}, E_{(2)})$, \textbf{HAGNE} first generates corresponding graph embeddings $\mathbf{h}_{G_{(1)}}$ and $\mathbf{h}_{G_{(2)}}$ in a hierarchical attentional style and then fuses the two graph embeddings via Similarity Learning (\textbf{SL}) and outputs a similarity score $Sim(G_{(1)},G_{(2)})$. In this way, the representation of the graph and similarity metric can be learned in a joint way, such that the effective graph similarity matching can be performed. During the malware detection stage, the distance between an unknown program and an existing benign program will be maximized in the mapped embedding space under the learned similarity metric.

\nop{To address the challenges introduced in \Cref{sec:intro}, we propose a malicious program detection framework, \toole, with three key modules: Invariant Graph Modeling (IGM), Atentional Heterogeneous Graph Neural Network (AHGNN), and Learn to Detect via Siamese Network (LDSN) as illustrated in \Cref{fig:Architecture}. The IGM module models the system event data as a heterogeneous invariant graph, which can capture the program's behavior profile. The AHGNN module extracts the program embeddings from the constructed heterogeneous invariant graphs. Finally, the LDSN module aims to train the AHGNN to be more distinguishable between an unknown program and the existing benign programs.}

\subsection{Invariant Graph Modeling}
\label{subsec:graph}

Information systems often generate a large volume of system-event data (\textit{i.e.}, the interaction between a pair of system entities\nop{such as processes, files, and Internet sockets}). In a typical enterprise environment, the amount of data collected from a single computer system can easily reach one gigabyte after monitoring process interactions for one hour.

Learning a representation over such massive data is prohibitively expensive in terms of both time and space. Recently, a very promising means for studying complex systems has emerged through the concept of invariant graph~\cite{cheng2016ranking,LuoCTSLCY18}. Such invariant graph focuses on discovering stable and significant dependencies between pairs of entities that are monitored through surveillance data recordings, so as to profile the system status and perform subsequent reasoning. 

Following the idea of the invariant graph, we model the system event data as a heterogeneous graph between different system entities (\textit{e.g.}, processes, files, and Internet sockets). The edges indicate the causal dependencies including a process accessing a file, a process forking another process, and a process connecting to an Internet socket. Formally, given the event data $U$ across several machines within a time window (\textit{e.g.}, one day), each target program can be a heterogeneous graph $G = (V,E)$, in which $V$ denotes a set of nodes. Each node represents an entity of three possible types: process (P), file (F), and INETSocket (I), namely $V = P \cup F \cup I$.
$E$ denotes a set of edges (dependencies) $(v_s,v_d, r)$ between the source entity $v_s$ and destination entity $v_d$ with relation $r$. We consider three types of relations: (1) a process forking another process ($P \rightarrow P$), (2) a process accessing a file ($P \rightarrow F$), and (3) a process connecting to an Internet socket ($P \rightarrow I$). Each graph is associated with an adjacency matrix $A$. With the help of the invariant graph modeling, we can obtain a global program-dependency profile.

\subsection{Hierarchical Attentional Graph Neural Encoder}
The constructed invariant graph is heterogeneous with multiple types of entities and relations.
Thus, it is difficult to directly apply the traditional homogeneous Graph Neural Networks (such as GCN and GraphSage) to learn the graph representation. To address this problem, we propose a Hierarchical Attentional Graph Neural Encoder (\textbf{HAGNE}) 
to learn the program representation as a graph embedding through an attentional architecture by considering the node-wise, layer-wise, and path-wise context importance.
More specifically, we first propose a \textbf{Heterogeneity-aware Contextual Search} (Step (B1) in Figure~\ref{fig:Architecture}) to find the path-relevant sets of neighbors under the guide of the meta-paths \cite{sun2011pathsim}. Then, we introduce a \textbf{Node-wise Attentional Neural Aggregator} (Step (B2) in Figure~\ref{fig:Architecture}) to generate node embeddings by selectively aggregating the entities in each path-relevant neighbor set based on random walk scores. Next, we design a \textbf{Layer-wise Dense-connected Neural Aggregator} (Step (B3) in Figure~\ref{fig:Architecture}) to aggregate the node embeddings generated from different layers towards a dense-connected node embedding. Finally, we develop a \textbf{Path-wise Attentional Neural Aggregator} (Step (B4) in Figure~\ref{fig:Architecture}) to learn the attentional weights for different meta-paths and compute the graph embedding from the Layer-wise Dense-connected Aggregator.

\subsubsection{Heterogeneity-aware Contextual Search}
As the first step of the aggregation layer, traditional GNNs would search for all one-hop neighbors for a target node. Since our invariant graph is heterogeneous, simply aggregating these neighbors cannot capture the semantic and structural correlations among different types of entities. To address this issue, we propose a meta-path~\cite{sun2011pathsim} based contextual search. A meta-path is a path that connects different entity types via a sequence of relations in a heterogeneous graph. In a computer system, a meta-path could be: a process forking another process ($P \rightarrow P$), two processes accessing the same file ($P \leftarrow F \rightarrow P$), or two processes opening the same internet socket ($P \leftarrow I \rightarrow P$) with each one defining a unique relationship between two programs. From the invariant graph $G$, a set of meta-paths  $\mathcal{M}=\{M_{1},M_{2},...M_{|\mathcal{M}|}\}$ can be generated with each representing a unique multi-hop relationship between two programs. For each meta-path $M_i$ where $i \in \{1,2,...,|\mathcal{M}|\}$, we define the path-relevant neighbor set $\mathcal{N}_v^i$ of node $v$:
\begin{equation}
\vspace{-1pt}
\mathcal{N}_v^i= \{u|(u,v)\in M_{i}, (v,u)\in M_{i}\}
\end{equation}
where $u$ is a reachable neighbor of $v$ via the meta-path $M_i$.

\subsubsection{Node-wise Attentional Neural Aggregator}
After constructing the path-relevant neighbor set $\mathcal{N}_v^i$, we are able to leverage these contexts via neighbourhood aggregation. However, due to noisy neighbors, different neighboring nodes may have different impacts on the target node. Hence, it is unreasonable to treat all neighbors equally. 
To address this issue, we propose a node-wise attentional neural aggregator to compute an attentional weight for each node in the path-relevant neighbor set $\mathcal{N}_v^i$. This module is based on random walk with restarts (RWR)~\cite{Tong2006}, a particularly efficient algorithm for computing the relevance scores between pairs of nodes in a homogeneous graph. 
We extend the RWR to a heterogeneous graph, such that the walker starts at the target program node $v$, and at each step it only moves to one of its neighboring nodes in $\mathcal{N}_v^i$, 
instead of to all linked nodes without considering semantics. After the random walk finishes, each visited neighbor will receive a visiting count. We compute the $L_1$ normalization of the visiting count and use it as the node-wise attentional weight. Formally, for $\mathcal{N}_{v}^{i}= \{u_{1}^{i}, ..., u_{n}^{i} \}$, the attentional weights are $\boldsymbol{\alpha}_{(v)(:)}^{i}= [\alpha_{(v)(1)}^{i}, ..., \alpha_{(v)(n)}^{i}]$, where $\alpha_{(v)(j)}^{i}$ is the weight of $u_{j}^{i}$. Then, the program representation can be computed via a neural aggregation function $AGG_{node}$ by
\begin{equation}
\footnotesize
\begin{aligned}
 &\mathbf{h}_v^{(i)(k)} = AGG_{node}(\mathbf{h}_v^{(i)(k-1)}, \{\mathbf{h}_u^{(i)(k-1)}\}_{u \in \mathcal{N}_v^i})\\
                   & = MLP^{(k)}((1+\epsilon^{(k)}) \mathbf{h}_v^{(i)(k-1)} + \sum_{u \in \mathcal{N}_v^i} \boldsymbol{\alpha}_{(u)(:)}^i \mathbf{h}_u^{(i)(k-1)})
\end{aligned}
\end{equation}
where $k\in\{1,2,...K\}$ denotes the index of the layer, $\mathbf{h}_v^{(i)(k)}$ is the feature vector of program $v$ for meta-path $M_i$ at the $k$-th layer, and $\epsilon^{(k)}$ is a trainable parameter that quantifies the trade-off between the previous layer representation and the aggregated contextual representation. $\mathbf{h}_u^{(i)(0)}$ is initialized by $X_v$. After getting the aggregated representation, a multi-layer perceptron (MLP) is applied to transform the aggregated representation to a hidden nonlinear space. Through the development of Node-wise Attentional Neural Aggregator, we are able to leverage the contextual information, meanwhile considering the different importance of the neighbors.

\subsubsection{Layer-wise Dense-connected Neural Aggregator}
To aggregate the information from a wider range of neighbors, 
one simple way is to stack multiple node-wise neural aggregators. 
However, the performance of a GNN model often cannot get improved, because by adding more layers, it is easy to propagate the noisy information from an exponentially increasing number of neighbors in a deep layer. Recently, inspired by residual network, a \textbf{skip-connection} method has been proposed.
But, even with skip-connections, GCNs with more layers do not perform as well as the $2$-layer GCN on many graph data \cite{kipf2016semi}.
To address the limitations of existing work, we propose a Layer-wise Dense-connected Aggregator as inspired by the \textsc{DenseNet}~\cite{huang2017densely}. More specifically, our layer-wise aggregator leverages all the intermediate representations, with each capturing a subgraph structure. All the intermediate representations are aggregated in a concatenation way followed by a MLP, such that the resulted embedding can adaptively select different subgraph structures. Formally, the Layer-wise Dense-connected Neural Aggregator $AGG_{layer}()$ can be constructed as follows:
\begin{align}
\mathbf{h}_{v}^{(i)(K+1)} & = AGG_{layer}(\mathbf{h}_v^{(i)(0)}, \mathbf{h}_v^{(i)(1)},... \mathbf{h}_v^{(i)(K)})\\
                & = MLP([\mathbf{h}_v^{(i)(0)}; \mathbf{h}_v^{(i)(1)};... \mathbf{h}_v^{(i)(K)}])
\end{align}
where $[\cdot; \cdot]$ represents the feature concatenation operation.

\subsubsection{Path-wise Attentional Neural Aggregator}
After the node-wise and layer-wise aggregators, different embeddings corresponding to different meta-paths are generated. However, different meta-paths should not be treated equally. For example, Ransomware is usually very active in accessing files, but it barely forks another process or opens an internet socket, while VPNFilter is generally very active in opening the internet socket, but it barely accesses a file or forks another process. Therefore, we need to treat different meta-paths differently. To address this issue, we propose a Path-wise Attentional Neural Aggregator to aggregate the embeddings generated from different path-relevant neighbor sets. Our path-wise aggregator can learn the attentional weights for different meta-paths automatically.
Formally, given the program embedding $\mathbf{h}_v^{(i)(K+1)}$ corresponding to each meta-path $M_i$, we define the path-wise attentional weight as follows: 
\begin{equation}
\beta_{i}=\frac{\exp (\sigma(\mathbf{b}[\mathbf{W}_b \mathbf{h}_v^{(i)(K+1)}||\mathbf{W}_b \mathbf{h}_v^{(j)(K+1)}]))}{\sum_{j'\in |\mathcal{M}|} \exp (\sigma (\mathbf{b}[\mathbf{W}_b \mathbf{h}_v^{(i)(K+1)}||\mathbf{W}_b \mathbf{h}_v^{(j')(K+1)}]))}
\label{eq:att}
\end{equation}
where $\mathbf{h}_v^{(i)(K+1)}$ is the embedding corresponding to the target meta-path $M_i$, $\mathbf{h}_v^{(j)(K+1)}$ denotes the embedding corresponding to the other meta-path $M_j$, $\mathbf{b}$ denotes a trainable attention vector, $\mathbf{W}_{b}$ denotes a trainable weight matrix, which maps the input features to the hidden space, $||$ denotes the concatenation operation, and $\sigma$ denotes the nonlinear gating function. We formulate a feed-forward neural network, which computes the correlation between one path-relevant neighbor set and other path-relevant neighbor sets. This correlation is normalized by a Softmax function. Let $ATT()$ represent Eq.(\ref{eq:att}). The joint representation for all the meta-paths can be represented as follows:
\vspace{-5pt}
\begin{equation}
\mathbf{h}_{G}=AGG_{path}=\sum_{i=1}^{|\mathcal{M}|} ATT(\mathbf{h}_v^{(i)(K+1)}) \mathbf{h}_v^{(i)(K+1)}
\end{equation}
The Path-wise Attentional Neural Aggregator allows us to better infer the importance of different meta-paths by leveraging their correlations and learn a path-aware representation.

\subsection{Similarity Learning}
In order to perform effective graph matching, we propose Similarity Learning (\textbf{SL}), a Siamese
Network based learning model to train the parameters of the Hierarchical Attentional Graph Neural Encoder (\textbf{HAGNE}). Siamese Networks \cite{zagoruyko2015learning} are neural networks containing two or more identical subnetwork components, which have been shown to be a powerful way in distinguishing similar and dissimilar objects. Here, we employ the \textbf{SL} to learn similarity metric and program graph representation jointly for better graph matching between the pair of unknown program and known benign program as shown in Figure~\ref{fig:Architecture}.

More specifically,
our Siamese Network consists of two identical \textbf{HAGNE}s to compute the program graph representation independently. 
Each \textbf{HAGNE} takes a program graph snapshot as the input and outputs the corresponding embedding $\mathbf{h}_G$. A neural network is then used to fuse the two embeddings. The final output is the similarity score of the two program embeddings. 
During the training, $P$ pairs of program graph snapshots $(G_{i(1)},G_{i(2)}), i \in \{1,2,...P\}$ are collected with corresponding ground truth pairing information $y_i \in \{+1, -1\}$. If the pair of graph snapshots belong to the same program, the ground truth label is $y_i=+1$, otherwise its ground truth label is $y_i=-1$.
For each pair of program snapshots, a cosine score function is used to measure the similarity of the two program embeddings and the output of the Similarity Learning is defined as follows:
\begin{align}
Sim(G_{i(1)},G_{i(2)})
&=cos((\mathbf{h}_{G_{i(1)}},\mathbf{h}_{G_{i(2)}}))\\
&=\frac{\mathbf{h}_{G_{i(1)}} \cdot \mathbf{h}_{G_{i(2)}}}{||\mathbf{h}_{G_{i(1)}}|| \cdot ||\mathbf{h}_{G_{i(2)}}||}
\end{align}
Correspondingly, our objective function can be formulated as:
\begin{equation}
\ell=\sum_{i=1}^{P}(Sim(G_{i(1)},G_{i(2)})-y_i)^2
\end{equation}
We optimize this objective with Adam optimizer~\cite{kingma2014adam}.
With the help of the Similarity Learning, we can learn the parameters that keep similar embeddings closer while pushing dissimilar embeddings apart by directly optimizing the embedding distance. 

Since we directly optimize the distance between the two programs, this model can be used to perform unknown malware detection. Given the snapshot of an unknown program, we first construct its corresponding program invariant graph and then feed it to the \textbf{HAGNE} to generate the program embedding. After that, we compute the cosine distance scores between the embedding of the unknown program and the ones of the existing programs in the database. If an existing program has more than one embedding generated from multiple graph snapshots, we will only report the highest similarity score with regard to the unknown program. Finally, we rank all the similarity scores. If the highest similarity score among all the existing programs is below our threshold, an alert will be triggered. Otherwise, the top-$k$ most similar programs will be reported.



\section{Experiments}
\subsection{Experiment Setup}
\subsubsection{Data}
We collect a $20$-week period of data from a real enterprise network composed of $109$ hosts ($87$ Windows hosts and $22$ Linux hosts). The sheer size of the data set is around three terabytes. We consider three different types of system events as defined in \Cref{subsec:graph}. Each entity is associated with a set of attributes, and each process has an executable name as its identifier (ID). In total, there are about $300$ million event records, with about $2,000$ processes, $600,000$ files, and $18,000$ Internet sockets. Based on the system event data, we construct a program invariant graph per program per day.

\subsubsection{Baselines}
We compare \tool with the following typical and state-of-the-art algorithms:
\begin{itemize}
\item \textit{LR} and \textit{SVM}: LR and SVM represent the Logistic regression and Linear Support Vector Machine, respectively. They are two typical classification methods. We extract raw features from each process as the input, including the connectivity features and the graph statistic features. 

\item \textit{MLP} \cite{pal1992multilayer}: Multi-layer Perceptron (MLP) is a deep neural network based classification model with multiple non-linear layers between the input and the output layers. It is a special case of GNN without considering the aggregation operation if we define the propagation layer as an identity matrix. 
\end{itemize}

Since the proposed \tool is based on GNN, we also compare it with two popular GNN variants: GCN \cite{kipf2016semi} and  GraphSage \cite{hamilton2017inductive}. 

\nop{\begin{itemize}
\item \textit{GCN}:  is graph convolutional based semi-supervised classification algorithm. It is a special case of our contextual graph encoder. It consider all the 1-hop neighbourhoods of target entity as the receptive field in each graph encoder and extract the deeper relational feature through stacked multiple convolutional layer. It focus on homogeneous network.

\item \textit{GraphSage}: GraphSage \cite{hamilton2017inductive} is another convolutional-like graph neural network. It is also a special case of our contextual graph encoder. Its receptive field is constructed based on the neighborhood sampled using uniform distribution. It also focus on homogeneous network.
\end{itemize}}

\subsubsection{Evaluation Metrics}
Similar to \cite{Das:2007}, we evaluate the performance of different methods using accuracy (ACC), F-1 score, and AUC score.

\subsection{Fake Program Detection}
\label{subsec:fake}
Our first research question focuses on the accuracy of \tool detecting fake programs. Here, we define a fake program as the one that uses the ID of another program. It is a common method for adversaries to hide their attacks.

To simulate the execution of fake programs on a large scale, in our testing dataset (\textit{i.e.}, data from the seventh week), we manually seed $1,000$ fake programs. To do so, before feeding the monitoring data to \toole, we randomly replace the ID of a known program to the ID of another known program. This process simulates an adversary who wants to hide the use of a benign system tool in his/her attacks.

\begin{figure}[htb]
\centering
\includegraphics[width = 0.8\linewidth]{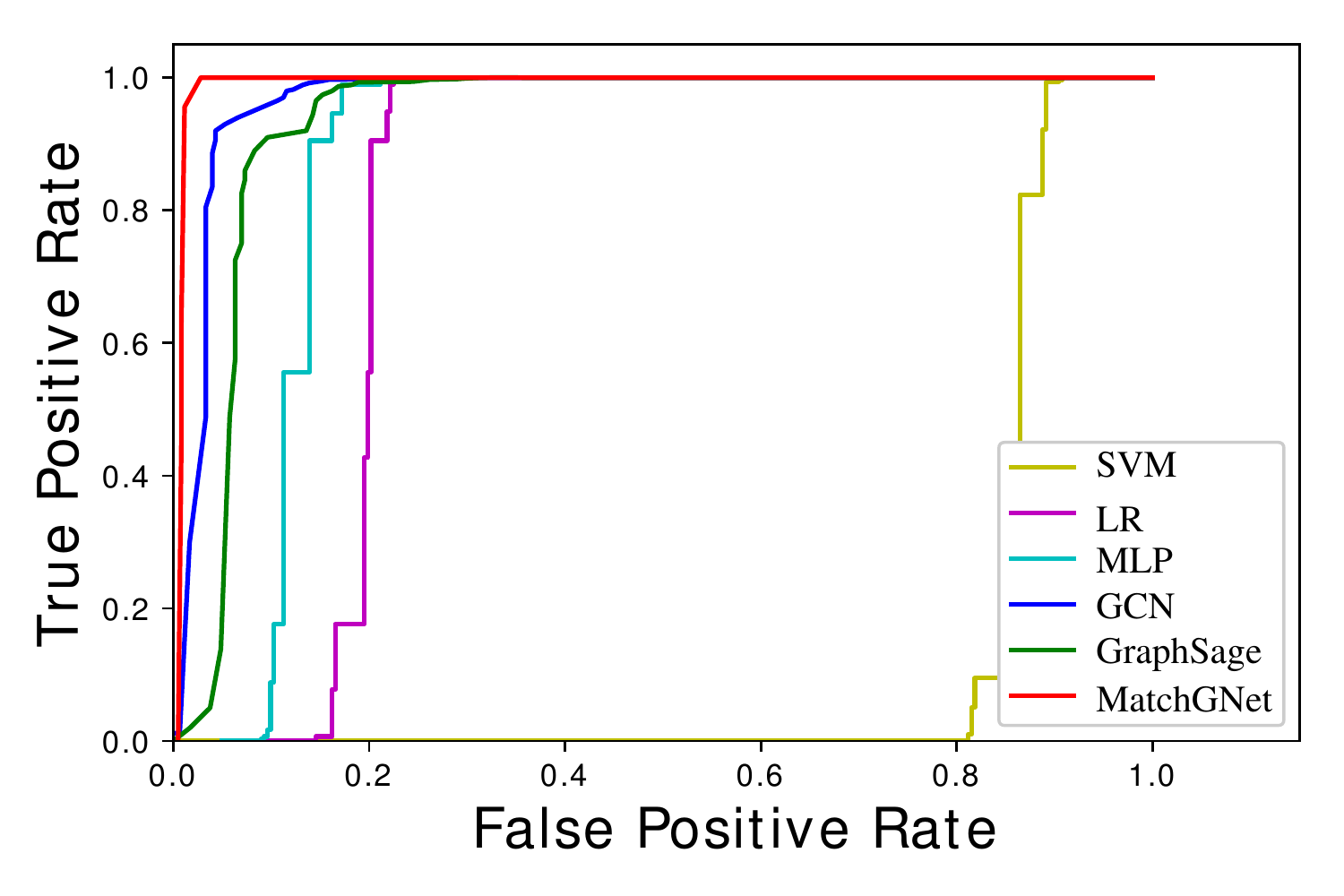}
\caption{ROC curves on detecting fake programs.}
\label{fig:DFP}
\end{figure}
\begin{table*}[htp]
\centering
\caption{Performance on fake program detection.}
\label{tab:sumfake}

\begin{tabular}{lllllll}
\hline
Method    & SVM   & LR    & MLP   &GCN &GraphSage &MatchGNet \\ \hline
AUC       & 51.11 & 82.79 & 85.11 &94.53 &91.72 &\textbf{98.51}  \\
F-1       & 66.07 & 83.05 & 86.12 &93.63 &90.02 & \textbf{98.49}  \\
ACC & 50.45 & 80.66 & 86.82 & 94.87& 92.03 &\textbf{97.03}  \\\hline
\end{tabular}

\end{table*}
\begin{table*}[htp]
\centering
\caption{Performance on unknown program detection.}
\label{tab:sumunknown}


\begin{tabular}{lllllll}
\hline
Method    & SVM   & LR    & MLP   &GCN &GraphSage &MatchGNet \\ \hline
AUC       & 51.11 & 80.63 & 83.69 & 92.05     & 90.38& \textbf{96.08}  \\
F-1       & 66.07 & 82.92 & 84.37 & 92.03     & 90.43& \textbf{95.55}  \\
ACC      & 50.01  & 81.98 & 85.00 & 92.24     & 91.50 &\textbf{96.53}  \\\hline
\end{tabular}

\end{table*}

In this task, each fake program has a claimed ID of a known program. Thus, we only need to check whether it is indeed the claimed program: if it matches the behavior pattern of the claimed program, the predicted label should be $-1$; otherwise, it should be $+1$. We use logistic regression as the classification model for all the methods.

The ROC curve is shown in \Cref{fig:DFP}. 
We also summarize the \textsc{AUC}, F-1 score and ACC of the six models in Table \ref{tab:sumfake}. From our experiments, the \textsc{AUC} of \tool is 99\%, which is at least 4\% higher than all other baseline models. In terms of ACC, \tool could achieve 47\%, 16\%, 10\%, 2\%, and 5\% higher than the SVM, LR, MLP, GCN, and GraphSage models. This means that while capturing all the fake program instances, \tool has less false positives than all other models. This result justifies our design decision of applying the invariant graph structure and Graph Neural Network model to capture the semantics of program instances.

\subsubsection{Hyper Parameter Selection} 
\begin{figure}[htb]
    \centering
    \begin{subfigure}[t]{0.48\linewidth}
        \centering
        \includegraphics[width=\linewidth]{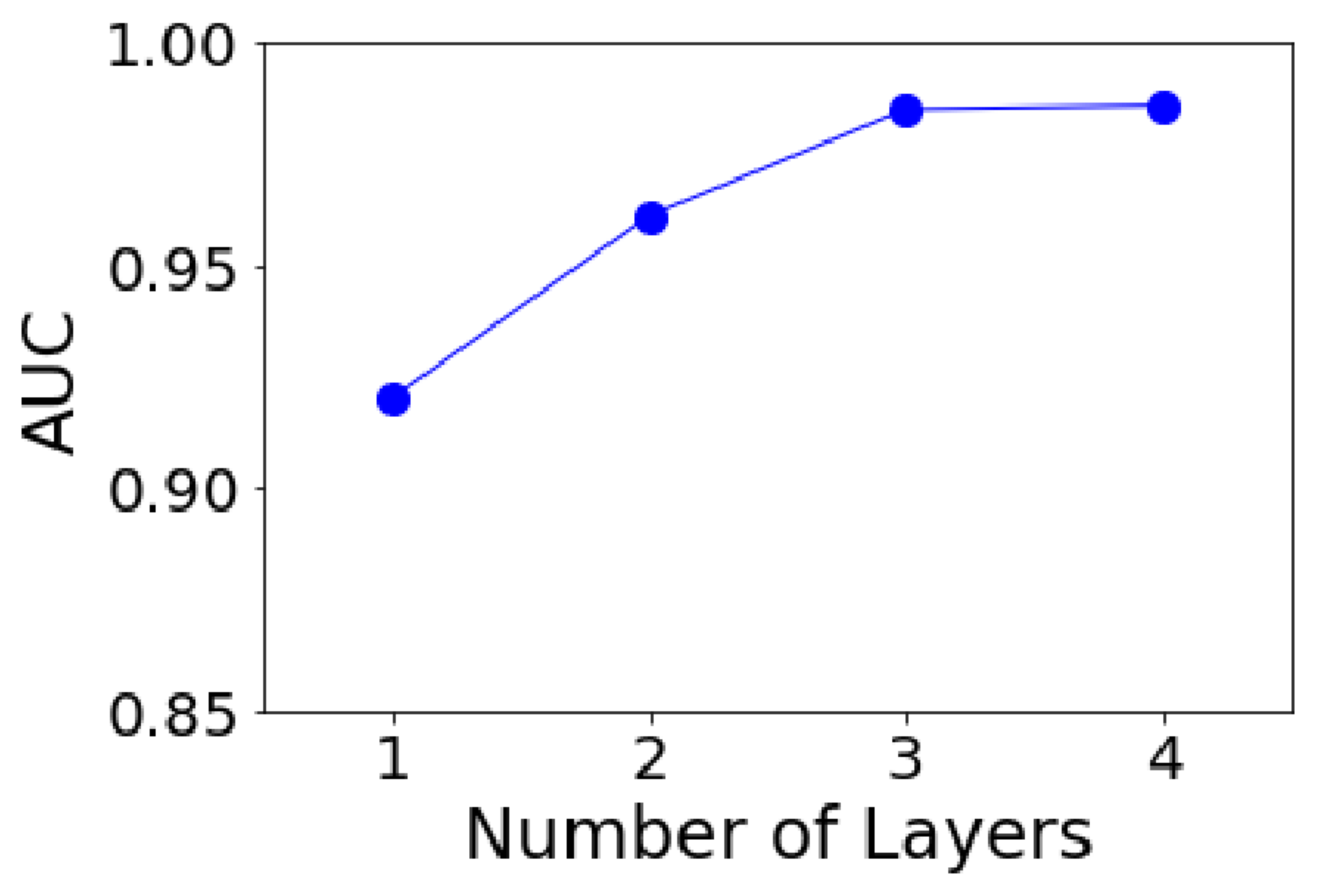}
  
        \label{fig:real1_TPR}
        \caption{AUC vs. the number of layers.}
    \end{subfigure}%
    ~ 
    \begin{subfigure}[t]{0.48\linewidth}
        \centering
        \includegraphics[width=\linewidth]{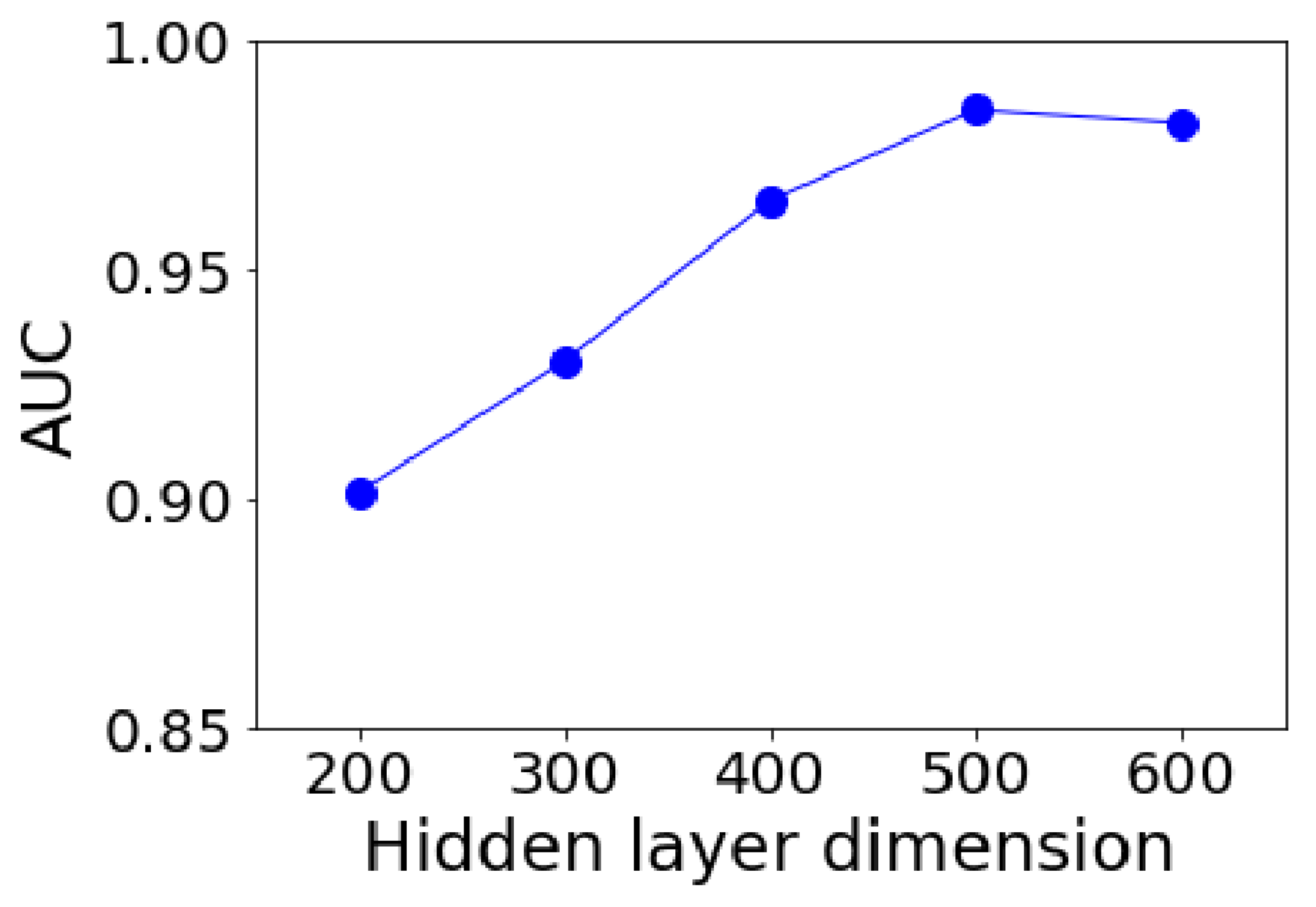}
       
        \label{fig:real1_TPR}
        \caption{AUC vs. the number of hidden neurons.}
    \end{subfigure}%

    \caption{Hyper parameters in fake program detection.}\label{fig:parameter}

\end{figure}

We evaluate the selection of hyper parameters of \tool with our validating data set (\textit{i.e.,} data from the sixth week). There are two hyper parameters in \toole: the number of layers, and the number of hidden neurons. We plot the result for the number of layers and the number of hidden neurons in ~\Cref{fig:parameter}. In these figures, the y-axis is the AUC value and the x-axis is the value of the hyper parameter.

We find that when \tool has 3 layers and 500 neurons, it reaches the maximal AUC. Larger hyper-parameter values may consume more resources but have little improvement on the AUC. Thus, we use the optimal hyper parameters as a part of the default model and apply them to the other parts of our experiments.

\subsection{Unknown Program Detection}
\label{subsec:unknown}
This experiment focuses on evaluating \tool on unknown program detection. To simulate unknown program instances, we split the programs in the training data equally into two sets, the known set and the unknown set. In our five weeks' training data, we exclude the programs in the unknown set and only train the model from the programs in the known set. Then, in our testing period, we use program instances from the unknown set as malicious programs. 

\begin{figure}[t!]
\centering
\includegraphics[width = 0.8\linewidth]{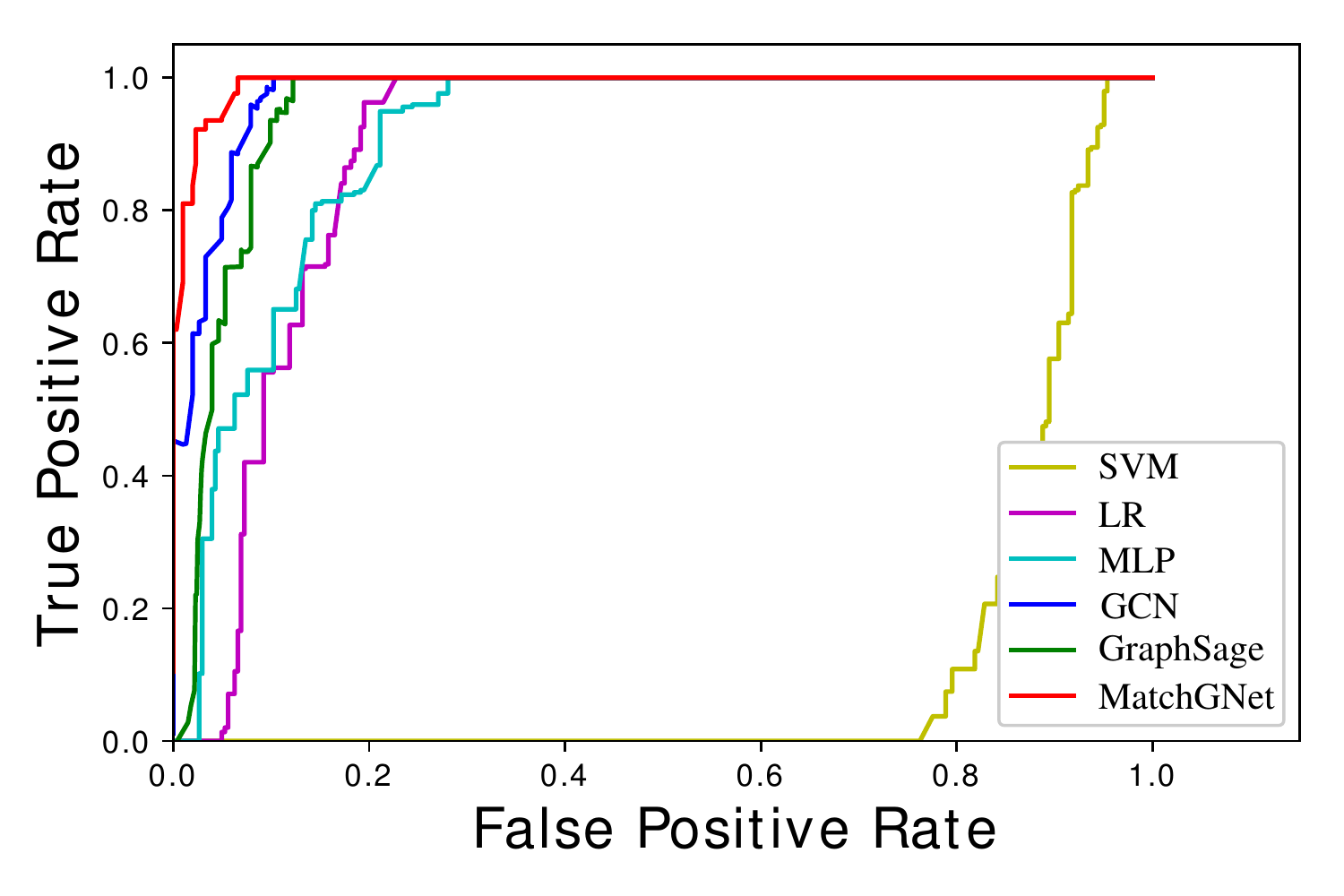}
\caption{ROC curves on detecting unknown programs.}
\label{fig:DUP}
\end{figure}
We plot the ROC curves in \Cref{fig:DUP} and report the AUC, F-1 score, and ACC of all six models in Table \ref{tab:sumunknown}. The AUC of \tool in detecting unknown program instances is at least 4\% higher than all other models. And in terms of ACC, \tool is 46\%, 14\%, 11\%, 4\%, and 5\% higher than the SVM, LR, MLP, GCN, and GraphSage models. This also proves that using the Attentional Graph Neural Network model could better capture the semantic level features of programs and may possibly generate less false positives.

\nop{We plot the ROC curves of \tool and the baselines in \Cref{fig:DUP}. In \Cref{tab:sumunknown}, we report the AUC of all six models. \Cref{tab:sumunknown} also contains the F-1 score and precision of all six models when the recall is 100\%. The AUC of \tool in detecting unknown program instances is 45\%, 9\%, and 11\% higher, respectively, than the SVM, LR, and MLP models. When the recall is 100\%, \tool has a precision 44\%, 15\%, and 13\% higher than the SVM, LR, and MLP models. In other words, without missing an unknown program, \tool could generate 44\%, 15\%, and 13\% less false positives than the SVM, LR, and MLP models. This also proves that using the Attentional Graph Neural Network model could better capture the semantic level features of programs and may possibly generate less false positives.}

\subsection{Malware Attack Detection}
\label{subsec:malware}

To evaluate the usefulness of \toole, we apply it to detect realistic malware attacks in enterprise environment. We randomly download $145$ malware from VirusTotal covering all the popular malware categories and create $9$ realistic attack cases from the literature and report including WannaCry, Genasom, Sorikrypt, Shinolocker, Puishing Email, ShellShock, Netcat Backdoor, Passing the Hash, and Trojan Attacks. 

We execute these attacks on our two testing machines and merge the provenance data from these two testing machines with the normal data of the $109$ hosts to the data stream of \toole. We set up \tool with the optimal hyper parameters in \Cref{subsec:fake}.

To evaluate the usefulness of \toole, we use the state-of-the-art entity-embedding based technique \cite{Chen:2016} as the baseline. 

Based on the experimental results, both \tool and the baseline method capture all the $154$ true alerts. However, during the same time period and with the same provenance data, \tool only generates $57$ false positives while the baseline generates $115$ false positives. Such a number indicates that \tool is possible to reduce nearly 50\% of the false positives. This reduction could mean a substantial saving in cyber-analysts' time in a large enterprise.

\section{Conclusion}
In this paper, we proposed \toole, a heterogeneous Graph Matching Network approach to detect the unknown malicious programs in information systems. \tool first models the program's execution behavior as a heterogeneous invariant graph. Based on the built program graphs, it learns the graph representation and similarity metric simultaneously to distinguish the benign program and malware.\nop{For any unknown testing program, \toole would also search for the existing program database and recommend the top-$k$ most similar programs, which would be very helpful for performing subsequent reasoning.} The evaluation results showed that our approach can accurately detect unexpected program instances. In particular, it can detect fake and unknown programs with an average accuracy of 97\% and 96\%, respectively. We further demonstrate our approach is promising by having 50\% less false positives than the state-of-the-art method in detecting malware attacks.

\section*{Acknowledgements}
Shen is supported in part by NSF through grants IIS-1526499, IIS-1763325, and CNS-1626432.

\bibliographystyle{named}
\bibliography{references}

\end{document}